\pdfoutput=1

\documentclass[useAMS,usenatbib]{mn2e}
\usepackage{graphicx,subfigure,amssymb,natbib}
\usepackage{amsmath}
\usepackage{amssymb}
\usepackage{epstopdf}
\usepackage{hyperref}
\usepackage{csquotes}

\title[Revised radio source count predictions]{9C spectral-index distributions and source-count estimates from 15 to 93~GHz --- a re-assessment}
\author[Waldram, Bolton, Riley,  Pooley]{E. M. Waldram$^{1}$, R. C. Bolton$^{1}$\thanks{E-mail: rosie@mrao.cam.ac.uk}, J. M. Riley$^{1}$, G. G. Pooley$^{1}$
\\ \\
$^{1}$Astrophysics Group, Cavendish Laboratory, J. J. Thomson Avenue, Cambridge CB3 0HE \\}
\begin{document}

\date{\today}

\pagerange{\pageref{firstpage}--\pageref{lastpage}} \pubyear{2018}

\maketitle

\label{firstpage}

\begin{abstract}
In an earlier paper (2007) we used follow-up observations of a sample of sources from the 9C~survey at 15.2~GHz to derive a set of spectral-index distributions up to a frequency of 90~GHz.  These were based on simultaneous measurements made at 15.2~GHz with the Ryle telescope and at 22 and 43~GHz with the Karl G. Jansky Very Large Array (VLA). We used these distributions to make empirical estimates of source counts at 22, 30, 43, 70 and 90~GHz. In a later paper (2013) we took data at 15.7~GHz from the Arcminute Microkelvin Imager (AMI) and data at 93.2~GHz from the Combined Array for Research in Millimetre-wave Astronomy (CARMA) and estimated the source count at 93.2~GHz. In this paper we re-examine the data used in both papers and now believe that the VLA flux densities we measured at 43~GHz were significantly in error, being on average only about 70\% of their correct values. Here we present strong evidence for this conclusion and discuss the effect on the source-count estimates made in the 2007 paper. The source-count prediction in the 2013 paper is also revised. We make comparisons with spectral-index distributions and source counts from other telescopes, in particular with a recent deep 95~GHz source count measured by the South Pole Telescope.

We investigate reasons for the problem of the low VLA 43-GHz values and find a number of possible contributory factors, but none is sufficient on its own to account for such a large deficit.   

\end{abstract}

\begin{keywords}
 surveys -- galaxies: evolution  -- cosmic microwave background -- radio continuum: general. 

\end{keywords}

\section{Introduction}

In two previous papers---\citealt{w07} (hereafter Waldram07) and \citealt{davies13} (hereafter Davies13)---we used follow-up observations of samples of sources from the 9C survey at 15.2~GHz to make empirical estimates of the source counts at higher radio frequencies. In the first, we used data at 15~GHz from the Ryle telescope and at 22 and 43~GHz from the Karl G. Jansky Very Large Array (VLA) to make predictions of the source counts at 22, 30, 43, 70 and 90~GHz.  In the second, we used data at 15.7~GHz from the Arcminute Microkelvin Imager (AMI) and at 93.2~GHz from the Combined Array for Research in Millimetre-wave Astronomy (CARMA) to estimate the count at 93.2~GHz. There proved to be a significant discrepancy between the results, in that the differential 93-GHz count from the second paper was approximately twice as high as the earlier 90-GHz prediction. Here we re-assess the data used in both papers. In particular, we conclude that the VLA 43-GHz values used in the first paper were systematically underestimated, being on average only about 70\% of their true values. We discuss the effect of this on the spectral index distributions and the source-count estimates.

Knowledge of the source counts at high radio frequencies is of considerable importance. Radio sources constitute a significant foreground contaminant in cosmic microwave background (CMB) observations whether for measurements of the CMB power spectrum or of the Sunyaev-Zel'dovich effect. In the analysis it is essential not only to subtract the brighter sources but also to take account of the cumulative effect of the fainter unsubtracted sources below the detection limit. (See, for example, \citealt{schneider12}.) Also, the counts are intrinsically very interesting in themselves for the understanding of radio-source evolution and comparison with current models (e.g. \citealt{toffolatti98}, \citealt{dezotti05}, \citealt{tucci11}).

Direct measurement of the source counts at high radio frequencies is notoriously difficult. (A summary of the available high-frequency counts was included in the review paper of \citealt{dezotti10}.) Blind  surveys at these frequencies can be excessively time-consuming, owing to the inevitably smaller fields of view and the decreasing flux densities of the majority of the sources. For high flux densities there are high-frequency counts---complete to 1~Jy---derived directly from surveys with both the WMAP satellite (\citealt{massardi09}) and the \emph{Planck} satellite (\citealt{planckXIII}). 

At the time of publication of Waldram07 and Davies13 there were no directly-measured high frequency counts at lower flux densities. It was therefore necessary to rely on indirect methods based on observed counts at somewhat lower frequencies. There were measured counts from deeper surveys at frequencies of 15~GHz (the 9C and 10C surveys: \citealt{w03}, \citealt{w10}, \citealt{franzen11}, \citealt{davies11}) and 20~GHz (the AT20G survey: \citealt{murphy10}, \citealt{massardi11}) and it was possible to use these, in conjunction with spectral index information from appropriate samples of sources, to make empirical estimates of the source counts at higher frequencies. This was the method used in our two papers, Waldram07 and Davies13, and also by \citealt{sadler08} (hereafter Sadler08) and \citealt{sajina11} (hereafter Sajina11).

Since the publication of Waldram07 and Davies13, there has been a deep directly measured count at 95~GHz, down to 11~mJy, published by Mocanu et al. from the South Pole Telescope (\citealt{mocanu13}) and this is clearly more reliable than any indirect estimate at that frequency.  However, we believe it is important to publish this re-assessment of our earlier work, since the predictions in the 2007 paper, over a range of frequencies, have been used by a number of authors, both in the estimation of foreground contamination in CMB observations (e.g \citealt{tucci12}, \citealt{osborne14}) and in comparisons of the counts with other predictions and models (e.g. \citealt{tucci11}).

\section{Outline of the paper}

As explained above, this revision of our earlier work was prompted by finding such a significant discrepancy between the results in Waldram07 and in Davies13. This paper consists predominantly of the re-assessment of the data used in Waldram07. However, in making a detailed comparison with Davies13 we found a problem with the calibration of the 15.7~GHz data used in that paper and have made appropriate corrections.

The paper is set out as follows. 

\begin{description}
\item[\textbf{Section 3:} ]
The samples used in Waldram07 and Davies13. Correction to the calibration of the 15.7-GHz data in Davies13. 
\item[\textbf{Section 4:} ]
Spectral index distributions: the Waldram07-Davies13 (W-D) sample.
\item[\textbf{Section 5:} ]
Spectral index distributions: comparison of Waldram07 with samples from other instruments. 
\item[\textbf{Section 6:} ] Re-calculation of the spectra in Waldram07 omitting the 43~GHz data.
\item[\textbf{Section 7:} ] Estimates of the short-fall in the 43~GHz flux densities. 
\item[\textbf{Section 8:} ] Re-assessment of the source-count estimates in Waldram07 and Davies13. 
\item[\textbf{Section 9:} ] Comparison of the counts with other results.
\item[\textbf{Section 10:} ]
Investigation of possible reasons for the problem of the low VLA 43-GHz values.
\item[\textbf{Conclusions} ]
\end{description}

Note: we define the spectral index $\alpha$ for flux density $S$ and frequency $\nu$ such that $S \propto \nu^{-\alpha}$.

\section{The samples}

\subsection{The Waldram07 sample}

The Waldram07 spectral index distributions were based on a sample of 110 sources, from the 9C survey, which were followed up in 2001 and 2002 with simultaneous multi-frequency observations with the VLA, at 1.4, 4.8, 22 and 43 GHz. Observations were also made (within a very few days of the VLA ones) with the Ryle telescope (RT) in Cambridge at 15.2 GHz. This work has been described in detail in Bolton et al. 2004 (Bolton04 hereafter), Bolton et al. 2006a and Bolton et al. 2006b. 

\subsection{The Davies13 sample}

The Davies13 sample was also a subset from Bolton04 and consisted of 80 sources, 79 of which appeared in the Waldram07 sample. These were followed up with almost simultaneous observations at 15.7 GHz with AMI and 93.2~GHz with CARMA.

For the 79 sources in common between the two samples, we compared the RT 15.2 GHz flux densities in Waldram07 with the AMI 15.7 GHz flux densities in Davies13. There proved to be a systematic offset, with a median ratio for $S_{15.2}/S_{15.7}$ of 1.1, a value considerably too large to be due to the small difference in frequency. We have re-examined the AMI data. The sources were not bright enough for self-calibration but we have now found that the appropriate scale factor of 1.082, to correct for phase errors, was not applied. (The derivation of this factor is explained in Section~2.2 of the 10C paper, \citealt{davies11}). Also, an additional small scale factor should be applied to compensate for the small difference in frequency. Assuming a median spectral index of 0.5 (see again \citealt{davies11}), we estimate this to be 1.016.  The AMI 15.7-GHz data have therefore been multiplied by the product of two factors: 1.082 and 1.016. (The CARMA values are unaffected.) The corresponding correction to the predicted 93.2-GHz source count has been evaluated. (See Section~8.2.)

\section{Spectral index distributions: the Waldram07-Davies13 (W-D) sample}

We have used the 79 sources in common between the two samples to create the so-called `W-D sample'. This consists of the flux densities at 15.2, 22 and 43~GHz from Waldram07 and those at 93.2 GHz from Davies13. (Three of the sources were undetected at 93~GHz so for these we used the limiting flux density values from Davies13). 

\begin{figure}
\includegraphics[width=7.5cm, angle=270]{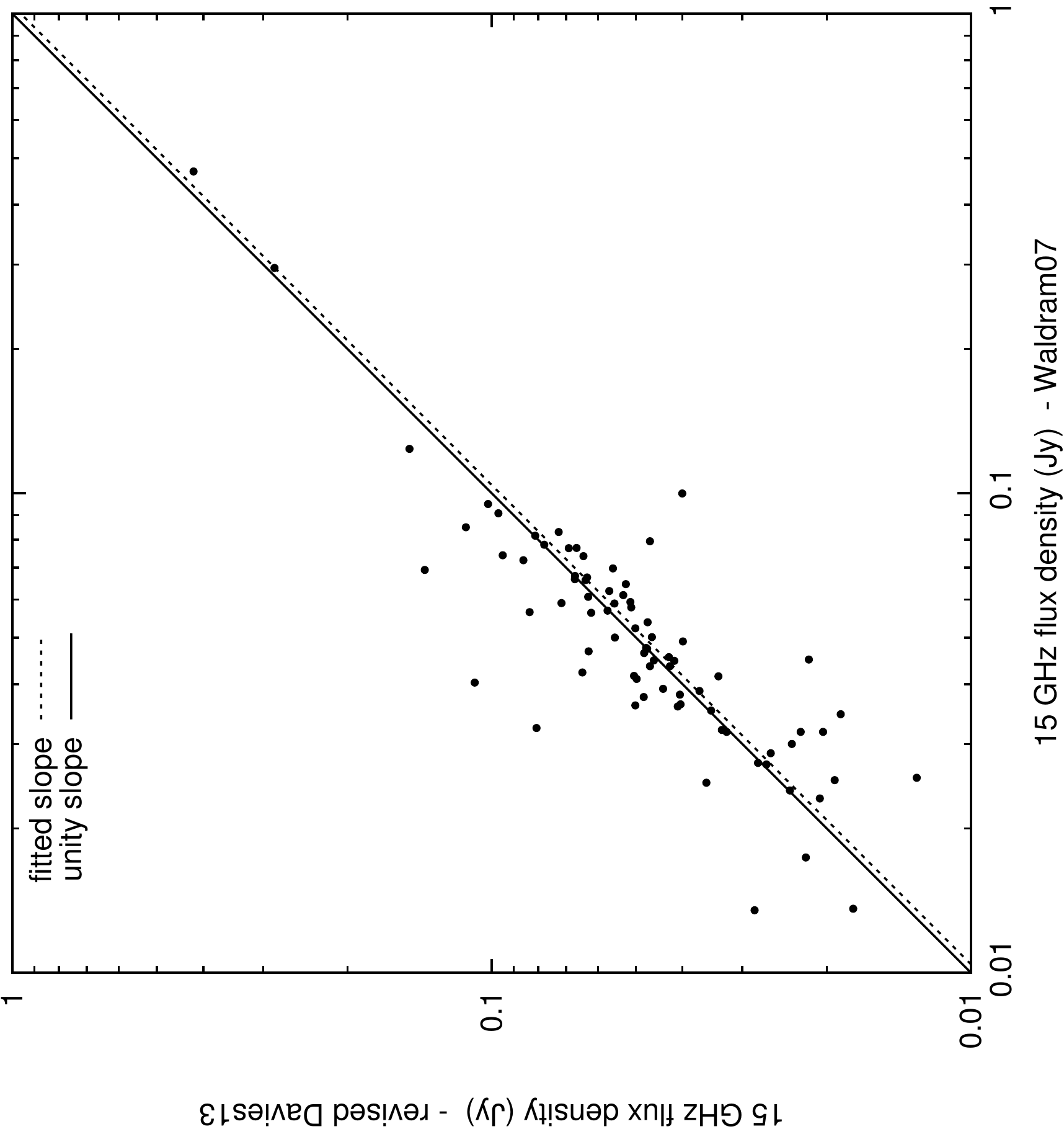}
\caption{15 GHz flux densities: revised Davies13 versus Waldram07}
\label{D13_vs_W07}
\end{figure}

The measurements from Waldram07 and Davies13 are not, of course, simultaneous.  Since, however, it is equally likely for an individual source to have increased or decreased in flux density, it should not be possible for variability to produce a significant systematic bias over the whole sample. We have plotted the 15~GHz values from the revised Davies13 sample versus those from the Waldram07 sample and find a slope close to unity ($0.96 \pm0.02$), as shown in Figure \ref{D13_vs_W07}.

We have examined the variation of the median spectral index with frequency in the W-D sample. (See Table \ref{SI_WD} and Figure \ref{spectra_compare}.) The spectral behaviour between  22, 43 and 93 GHz appears to be sufficiently unusual to throw doubt on the reliability of the flux-density measurements at 43 GHz. We find the median spectral index to be 1.14 between 22 and 43 GHz and 0.15 between  43 and 93 GHz, whereas between 22 and 93~GHz it is 0.6.

\section{Spectral index distributions: comparison of Waldram07 with samples from other instruments}

\subsection{Sadler08}

\begin{figure}
\includegraphics[width=7.5cm]{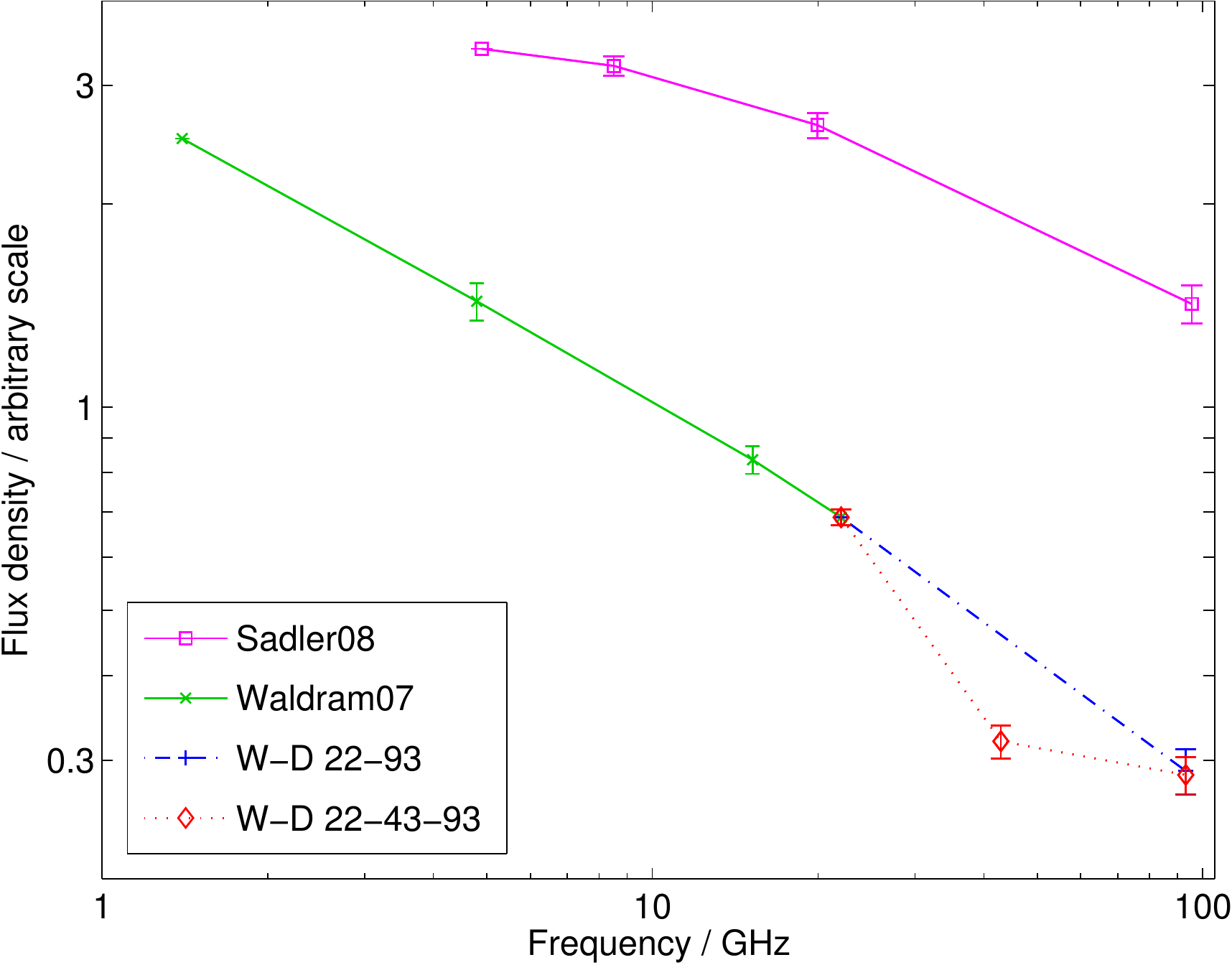}
\caption{Spectra of a `typical' source in the Waldram07 and W-D samples and the
AT20G sample (Sadler08).  The median spectral indices of these samples
in successive frequency ranges 
(Tables \ref{SI_WD} and \ref{SI_compare}) are used to indicate the
shape of the spectrum.  The flux density at the lowest frequency was
arbitrarily chosen and the median spectral indices used to predict,
successively, the flux density at higher frequencies.  The crosses
joined by the solid line show the spectrum from 1.4 to 22 GHz from the
Waldram07 data.  Above 22 GHz the spectral index data is from the W-D
sample, a sub-sample of the Waldram07 sample for which 93-GHz flux
densities have been measured.  The dot-dashed line shows the spectrum
deduced using the 22 and 93 GHz data only.  The diamonds joined by the
dotted line show the behaviour between 22 and 43 GHz and between 43
and 93 GHz; note the very abrupt change in slope at 43 GHz.  For
comparison, the squares joined by solid lines indicate the spectrum
for a typical source in the AT20G sample.}
\label{spectra_compare}
\end{figure}

There is also a marked discrepancy between the spectra in Waldram07 and those in Sadler08. (See Table \ref{SI_compare} and Figure \ref{spectra_compare}.) In the analysis in Waldram07 it was found that the median spectral index between 22 and 43 GHz was 1.06, compared with 0.53 for the median spectral index between 15 and 22 GHz.  In the absence of any data at a higher frequency it seemed plausible that the radio spectra should steepen in this way between 15 and 43 GHz.  Subsequently Sadler08, in their paper on 95-GHz observations of a sample of brighter sources from AT20G, also observed steepening in the spectra of their sources with increasing frequency---the median spectral index increased from 0.24 between 8 and 20 GHz to 0.39 between 20 and 95 GHz.  Although Sadler08 were looking across a greater frequency range, the steepening they observe is far less pronounced than that reported in Waldram07;  they comment that the differences between their results and those in Waldram07 could possibly be attributed to the different flux density ranges covered by the two samples---their sample is complete above 150 mJy at 20 GHz whilst the Waldram07 sources are much fainter, being complete above 25 mJy (with only four sources with flux densities greater than 150 mJy).

However, although the sources in AT20G are significantly brighter than those in Waldram07, the discrepancies are so striking that it seems unlikely that they are the result of these flux density differences. 

\subsection{Sajina11}

 A more recent follow-up at 4.9, 8.5, 22.5 and 43.3 GHz of a sample of 159 sources also from AT20G in Sajina11 shows that the median spectral index increases gently over this range, from 0.21 (4.9 to 8.5 GHz), to 0.37 (8.5 to 22.5 GHz) to 0.45 (22.5 to 43.3 GHz). This dataset has the advantage of having measured flux densities at 43.3 GHz and it clearly shows a trend in spectral index quite unlike that in Waldram07. (See Table \ref{SI_compare}.)

\begin{table}
\caption{The variation of median spectral index with frequency for the W-D sample}
\centering
\begin{tabular}{ccccc}
\hline \hline
\multicolumn{5}{c}{W-D sample} \\ \hline
& frequency && median & \\
& range && spectral & \\
& /GHz && index & \\
\hline
& 22--43 && $1.14^{*}$ & \\
& 43--93 && $0.15^{*}$ & \\
& 22--93 && 0.6 & \\ \hline 
\end{tabular}
\begin{tabular}{l}
* Suspect flux densities at 43 GHz \\
\end{tabular}
\label{SI_WD}
\end{table}

\begin{table*}
\caption{The variation of median spectral index with frequency for Waldram07, Sadler08 and Sajina11}
\begin{tabular}{cccccccccccc}
\hline \hline
\multicolumn{2}{c}{Waldram07}&&&&\multicolumn{2}{c}{Sadler08}&&&&\multicolumn{2}{c}{Sajina11} \\ \hline
frequency & median &&&& frequency & median &&&& frequency & median \\
range & spectral &&&& range & spectral &&&& range & spectral  \\
/GHz & index &&&& /GHz & index &&&& /GHz & index \\

\hline
1.4--4.8 & 0.45 &&&&&&&&&& \\
4.8--15.2 & 0.47 &&&& 5--8 & 0.11 &&&& 4.9--8.5 & 0.21 \\
15.2--22 & 0.53 &&&& 8--20 & 0.24 &&&& 8.5--22.5 & 0.37 \\
22--43 & $1.06^{*}$ &&&& 20--95 & 0.39 &&&& 22.5--43.3 & 0.45 \\
\hline \hline
\end{tabular}
\begin{tabular}{lccccccccccc}
* Suspect flux densities at 43 GHz &&&&&&&&&&& \\
\end{tabular}
\label{SI_compare}
\end{table*}

\section{Re-calculation of the spectra in Waldram07 omitting the 43-GHz data}

As discussed above, the behaviour shown in the W-D sample in itself is unexpected enough to call into question the 43-GHz flux densities.  Taken with the fact that it is also in such stark contrast to the results of Sadler08 and Sajina11, we are led to the conclusion that the Bolton04 43-GHz flux densities are significantly underestimated. To investigate further we have re-calculated the spectra in Waldram07 with the omission of the 43~GHz data. To do this, we have extended our fits to the spectra to the lower frequency of 4.8 GHz, giving three reasonably spaced data points at 4.8, 15 and 22 GHz for each source. We have estimated a constant spectral index for each source using simply the best-fitting (log)-linear fit to the 4.8, 15 and 22 GHz measurements, rather than attempting to fit a varying spectral index over this range.

\begin{figure}
\includegraphics[width=8cm]{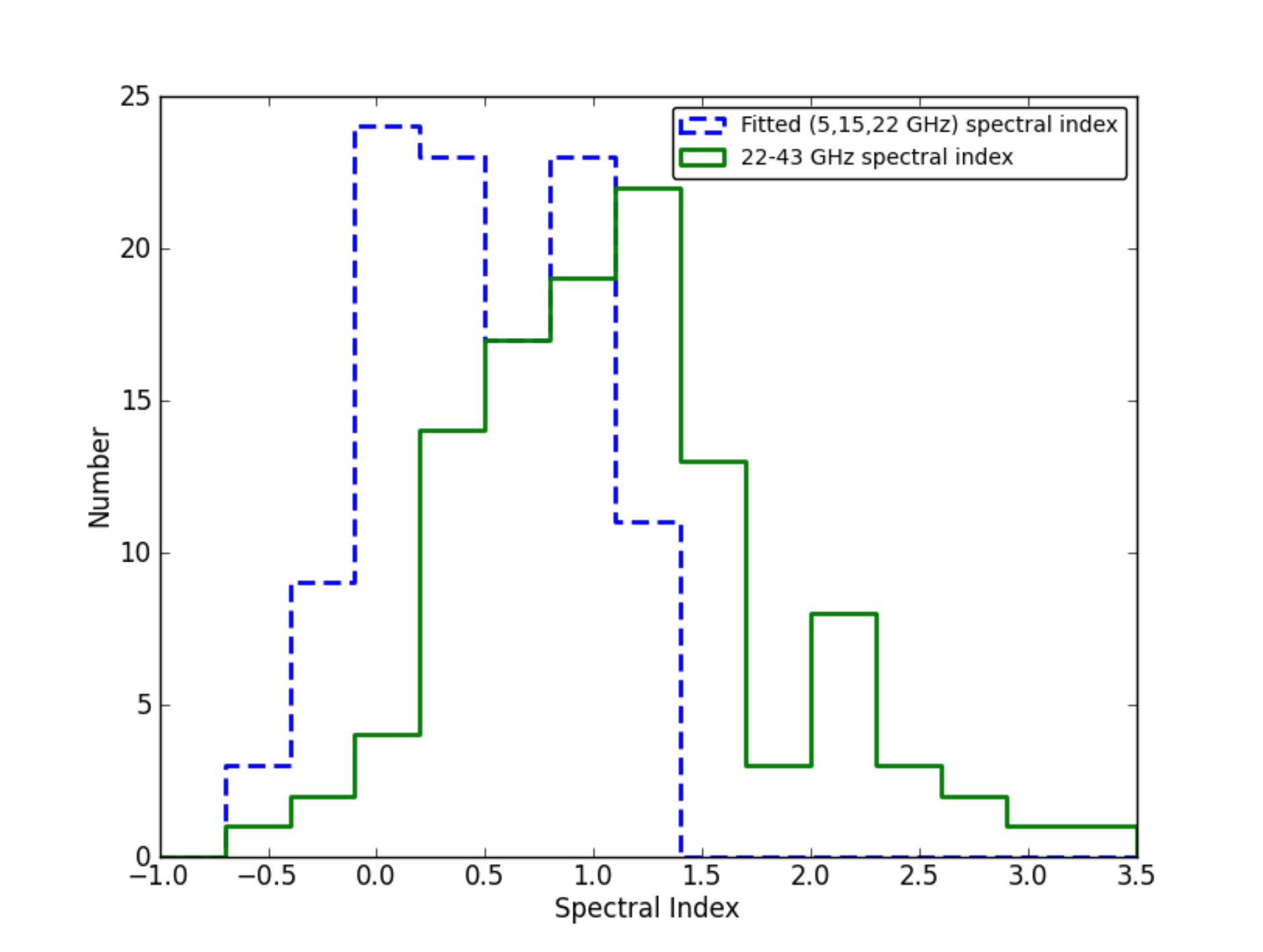}
\caption{ Histograms of spectral indices in the Waldram07 sample: distribution derived from fitting the 5, 15 and 22\,GHz values (blue dashed line) and from using the 22 and 43\,GHz values only (green solid line). }
\label{histos}
\end{figure}

In figure \ref{histos} we compare the distribution of the new spectral indices, derived from the fit to the 4.8, 15, 22-GHz flux densities ($\alpha_{5,15,22}$), with the distribution of spectral indices derived from the 22 and 43-GHz flux densities ($\alpha^{43}_{22}$), taking all 110 sources in the Waldram07 sample. There is clearly a large difference between these distributions. The new spectral indices ($\alpha_{5,15,22}$) are significantly lower than the 22-43\,GHz spectral indices ($\alpha^{43}_{22}$); the mean value of $\alpha_{5,15,22}$ is 0.47, as compared to 1.1 for $\alpha^{43}_{22}$.

\section{Estimates of the short-fall in the 43~GHz flux densities}

\subsection{Use of the W-D sample}

We have made a further investigation of the W-D sample, the 79 sources in Davies13 which also appeared in the Waldram07 sample. These are, in fact, the brighter sources from the original sample, so to test for any selection effect in their spectral properties, we calculated the median fitted spectral index ($\alpha_{5,15,22}$) for both the W-D sample (79 sources) and the Waldram07 sample (110 sources). We found values of 0.47 and 0.46 respectively, indicating that there is no significant spectral bias. We  then used the W-D sample to gain further insight into the problem of the VLA 43\,GHz flux density values.

We took the original 22-GHz flux densities (from Waldram07) and the CARMA 93-GHz values (from Davies13) and, using the spectral indices ($\alpha^{93}_{22}$), estimated the corresponding flux densities at 43\,GHz. Comparing these with the original 43-GHz measurements, we found a median value for the ratio of the measured to the predicted flux densities of 0.69. 

\subsection{Application of the $\alpha_{5,15,22}$ spectral indices to the Waldram07 sample}
We have also extrapolated the fitted spectral indices ($\alpha_{5,15,22}$) to predict the 43-GHz flux densities for all 110 sources in the Waldram07 sample and compared these with the original 43-GHz measurements. We found a median value for the ratio of the measured to the predicted flux densities of 0.66, which is consistent with the result from our 22-93\,GHz interpolation in the W-D sample described above.

\section{Re-assessment of the source-count estimates}

\subsection{The extrapolation method.}

In Waldram07 we showed that it is possible to use the known source count at one frequency $\nu_{1}$ to make an estimate of the count at a second frequency $\nu_{2}$, if we know the spectral index distribution between $\nu_{1}$ and $\nu_{2}$ of a complete sample of sources selected at frequency $\nu_{1}$. 

We showed that, if the differential count at $\nu_{1}$ has the form
\[n_{\nu_{1}}(S) = A_{\nu_{1}} S ^{-b}\]
where  $A_{\nu_{1}}$ and ${b}$ are constants,
then the count at $\nu_{2}$ can be written as
 \[n_{\nu_{2}}(S) = K A_{\nu_{1}} S ^{-b}\]
where K is a constant calculated from the spectral index distribution as follows.\\\\
For any one source we take the ratio $r$ of the flux densities at $\nu_{1}$ and $\nu_{2}$,  $r = S_{\nu_{1}}/S_{\nu_{2}} = \left({\nu_{1}}/{\nu_{2}}\right)^{-\alpha}$, where the spectral index is defined by $S\propto \nu^{-\alpha}$. Then for a sample of $m$ sources
 \[K = \frac{1}{m} \sum_{i=1}^{m} r_{i}^{1-{b}}.\]
(See \citealt{kellermann64} for a similar analysis.)\\\\
At 15~GHz the 9C differential source count (\citealt{w03}) is given by 
\begin{equation}
n(S) =  51\pm{3} \left(\frac{S}{\rm{Jy}}\right)^{-2.15} \:\: \rm{Jy}^{-1} sr^{-1},
\end{equation}
and the predicted count at another frequency $\nu$ becomes
\begin{equation}
n(S, \nu) = A(\nu) \left(\frac{S}{\rm{Jy}}\right)^{-2.15} \:\: \rm{Jy}^{-1} sr^{-1}.
\end{equation}
where $A(\nu) \approx K(\nu)\times51$

We use the error in the mean of the distribution $r_{i}^{1-{b}}$ to calulate the error in $A$.
\\

The method depends on two assumptions: first, that the sample provides a typical distribution of spectral indices and secondly, that this distribution is independent of flux density. We know that these assumptions are valid only over a limited range of flux density but, as shown in Waldram07, we can take this into account and estimate the range over which the revised counts are likely to be most reliable (see Figures \ref{compare_paco} and \ref{compare_spt}).

\subsection{Waldram07} 

In Waldram07 we calculated the predicted source counts at 30, 43, 70 and 90\,GHz on the basis of measurements at 15.2, 22 and 43~GHz. Since we now believe the 43-GHz values to have been significantly under-estimated, we expect the derived counts to be correspondingly too low. As a trial, in order to explore the extent of the short-fall, we have repeated the analysis in Waldram07, using simply the constant spectral indices $\alpha_{5,15,22}$ \emph{and omitting the 43~GHz points}.

In Table \ref{A Values} we present the $A(\nu)$ values at 43, 70 and 90\,GHz, obtained by our new fit, and compare these with the original Waldram07 values. As we have seen, our new analysis gives results with much lower spectral indices than the Waldram07 work. As a result our revised $A$ values are all higher, and the deviation between these and our original values increases with increasing frequency. However, these predictions are based on data only up to a frequency of 22~GHz. There may well be spectral steepening of a number of sources at higher frequencies which means that our revised $A$ values are likely to be somewhat over-estimated. (For brevity we refer to the revised and original counts  as `new Waldram07' and `old Waldram07'.) 

\subsection{Davies13}

We have also recalculated the predicted 93.2~GHz count in Davies13, using the same method. As explained in Section 3.2, the 15.7~GHz flux densities have been re-calibrated and increased by $\sim$ 10\%. This means that the revised 93.2~GHz count (or `new Davies13') is correspondingly lower than the original (or `old Davies13'), as shown in Table \ref{A Values}.

\begin{table}
\caption{The revised ($A_{\rm{new}}$)  values derived in this paper, compared with the original ($A_{\rm{old}}$) values in Waldram07 and Davies13, where: \newline $n(S)=A\times(S/\rm{Jy})^{-2.15}\rm{Jy}^{-1}sr^{-1}$}
\label{A Values} 
\begin{tabular}{lcccccc}
\hline
\hline
 Frequency & 43\,GHz & 70\,GHz & 90\,GHz & 93\,GHz \\
 \hline \\
Waldram07 data& \\
$A_{\rm{new}}$  & 34 $\pm$ 3 & 31 $\pm$ 4 & 30 $\pm$ 4 & 30 $\pm$ 4 \\ 
$A_{\rm{old}}$  & 22 $\pm$ 2 & 15 $\pm$ 2 & 13 $\pm$ 3 & - \\ \\
Davies13 data& \\
$A_{\rm{new}}$  & - & - & - & 23 $\pm$ 4   \\
$A_{\rm{old}}$  & - & - & - & 26 $\pm$ 4   \\
\hline
\hline
\end{tabular}
\end{table}

\section{Comparison of the counts with other results}

\subsection{The 40-GHz count from the PACO faint sample}

\begin{figure}
\includegraphics[width=7.5cm, angle=270]{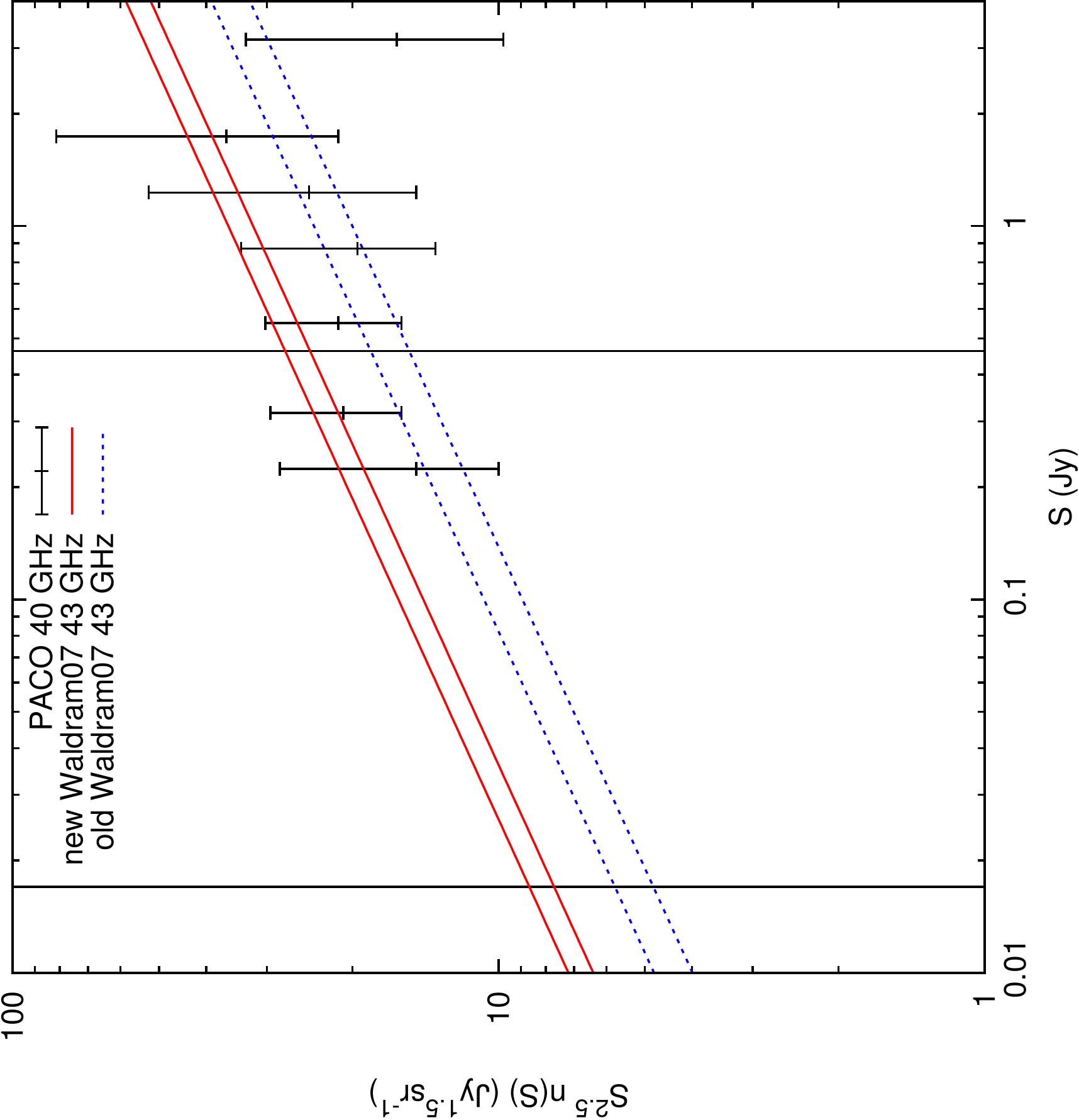}
\caption{ The new and old Waldram07 43-GHz counts compared with the 40-GHz count from the PACO faint sample, where the pairs of lines represent the estimated errors. The heavy vertical lines indicate the range of flux density over which the new count is likely to be most reliable.}
\label{compare_paco}
\end{figure}

We have compared our new and old 43-GHz counts with the 40-GHz count from the Planck-ATCA Coeval Observations (PACO) faint sample (\citealt{bonavera11}). This was a sample of 159 sources, complete to 200\,mJy at 20 GHz, which were observed at a range of frequencies between 4.5 and 40 GHz. The derived 40-GHz count is shown in Figure \ref{compare_paco} where we see that it is, in fact, consistent with either our new or old count. Owing to the large Poisson errors, it cannot usefully distinguish between them.

\subsection{The 95-GHz count from Sadler08}

Sadler et al. (Sadler08) took a flux-limited sample of sources from the AT20G survey and made simultaneous measurements of the 20- and 95-GHz flux densities with the Australia Telescope Compact Array. They then made their prediction from the spectral index distribution, together with the 20-GHz count, using a method similar to our own (\citealt{kellermann64}). We saw in section 5.1 that, in this sample, the steepening of the spectra with increasing frequency was found to be much less pronounced than that reported in Waldram07. We might, therefore, have expected the predicted 95-GHz count to have been significantly higher than the 90-GHz count predicted in Waldram07. However, as shown in Sadler08, the two counts were, in fact, in good agreement over the common flux-density range. 

At the time this was considered to be strong evidence for the validity of the predicted counts, since they were based on independent data from different telescopes. It was only later, when there proved to be such a significant discrepancy between the results in Davies13 and both Waldram07 and Sadler08, that there was further investigation. The question was explored in Davies13, where the conclusion was reached that the low 95-GHz count in Sadler08 arose from an under-estimate of the 20-GHz count on which it was based. In fact, the 20-GHz count was later revised upwards (\citealt{massardi11}) and in Davies13 we were able to show that this revised 20-GHz count was consistent with an appropriate extrapolation of the known 15-GHz count. \emph{Thus it appears that both the original Waldram07 90-GHz and the Sadler08 95-GHz counts were underestimated but for different reasons.}

\subsection{The 95-GHz count from the South Pole Telescope}

\begin{figure}
\includegraphics[width=7.5cm, angle=270]{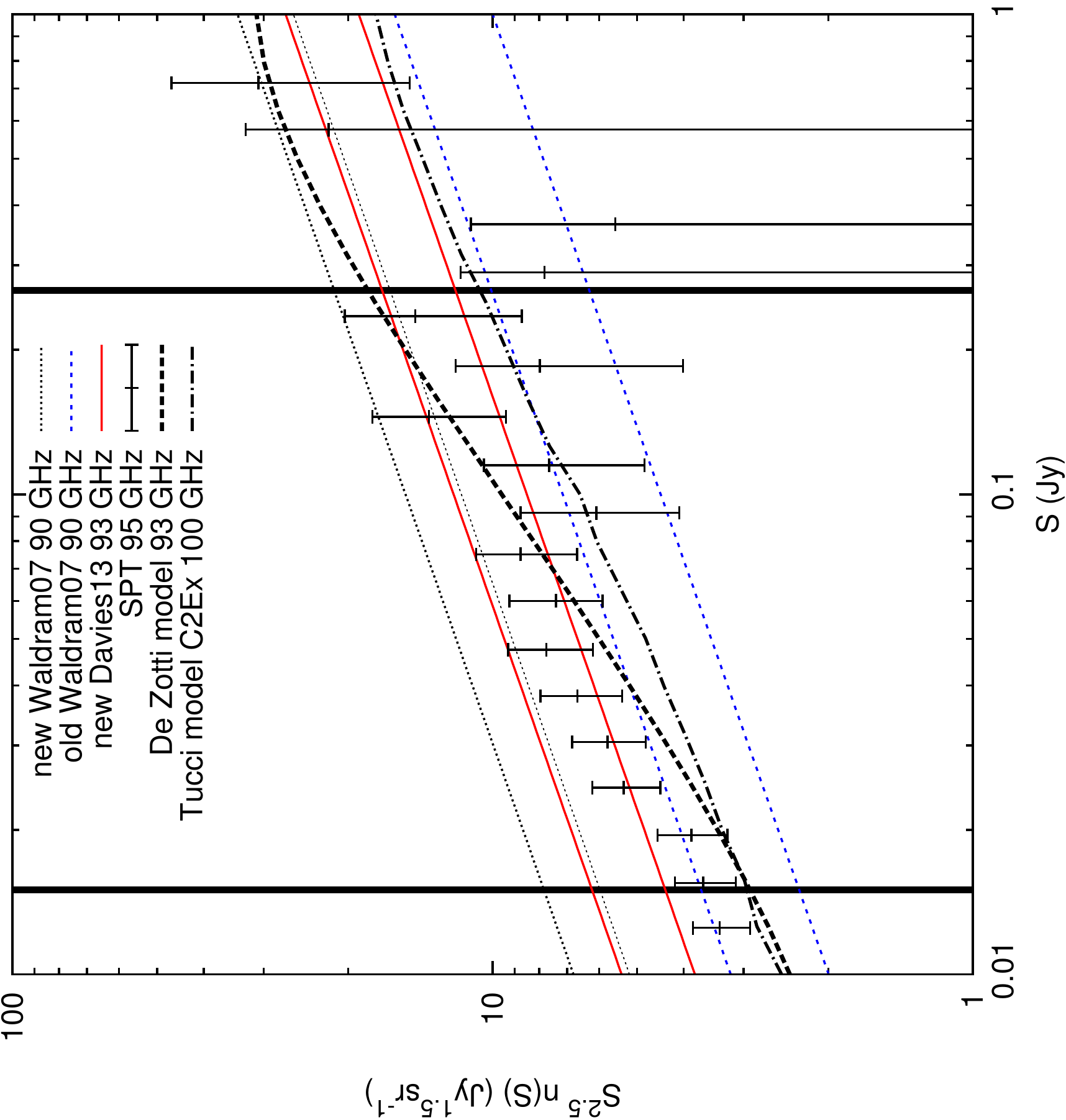}
\caption{Our predicted 90-GHz and 93-GHz counts compared with the 95-GHz count from the South Pole Telescope, where the pairs of lines represent the estimated errors. The heavy vertical lines indicate the range of flux density over which the new Davies count is likely to be most reliable. The range of reliability of the new Waldram07 count is estimated to be approximately 20--400 mJy.}
\label{compare_spt}
\end{figure}

As mentioned in the introduction, there is now a directly measured count at 95~GHz, down to 11~mJy, from the South Pole Telescope (SPT). This was unavailable at the time of publication of Waldram07 and Davies13 but we are now able compare our estimates with actual observations. In Figure \ref{compare_spt} we show the SPT 95-GHz count together with our predictions---the new and old Waldram07 90-GHz counts and the new Davies13 93-GHz count.  We see that, as expected, the old Waldram07 count is significantly too low; this is further evidence for the short fall in the VLA 43-GHz flux density values. The new Waldram07 90-GHz count is  higher than the SPT count; as explained in Section 8.1, this indicates that there must be some spectral steepening of some sources at frequencies above 22~GHz, even if it is not as extreme as originally assumed. The new Davies13 count is in good agreement with the SPT count over a range of approximately 20--250 mJy.

For comparison, we have included in this figure model counts from Gianfranco de Zotti and Marco Tucci (see \citealt{dezotti05}, \citealt{tucci11}). However, any discussion of the models is outside the scope of this paper.

\section{The 43-GHz VLA flux densities}

\subsection{Assessment of the problem}

In Section 7 we showed strong evidence that the 43-GHz flux densities used in Waldram07 were significantly in error, being, on average, only about 70\% of their correct values. In particular, we used the fitted spectral index ($\alpha_{5,15,22}$) to predict a 43-GHz flux density for all 110 sources and compared the predicted values with the measured 43-GHz values. We calculated the ratio of measured/predicted flux density for each source and found a mean scale factor ($F_{43}$) of $0.68 \pm 0.02$ (the median value was 0.66). This was in good agreement with  our investigation of the W-D sample, using the 22--93 GHz spectral indices, which indicated that the 43-GHz flux-density measurements in Waldram07 were low by a factor of 0.69.

\subsection{Analysis of the observations}

The 43-GHz flux densities used in Waldram07 were taken from the VLA measurements in Bolton04. To try to reveal the cause of the problem, we investigate here whether there are any observable trends in the 43\,GHz scale factor ($F_{43}$) with the different observing runs, since these will have undergone different calibration procedures and have used different antenna layouts. 

The Bolton04 data were taken during 2001 and 2002 (some years before the `EVLA' upgrade) in five separate observing runs at the VLA, with the array in a variety of different configurations. Of the 110 sources in the Waldram07 sample the majority  were observed in VLA runs numbered 1 (November 2001) and 5 (January 2002) in Bolton04. (The other VLA runs do not have sufficient numbers of sources from the sample of 110 to be statistically useful so we do not show results from these runs here.)

In Table \ref{scale1} first we take all sources that were part of runs 1 and 5 and in the Waldram07 sample and calculate the average $F_{43}$ value in each session. The derived scale factors for the two runs agree within the uncertainties. It is however possible that these results could be contaminated by variable objects and by sources with very non-linear fits (for which the predictions may be significantly wrong). To avoid this we repeat the analysis imposing a strict cut: including only sources with both steeply falling spectra (with $\alpha_{5,15,22} > 0.4$) {\it and} a correlation coefficient for the spectral index fit greater than 0.8. Selecting only falling spectrum sources should eliminate most variable sources (see \citealt{bolton06b}) and selecting sources with a well-fitted spectrum eliminates spectral curvature. For both VLA runs 1 and 5 the results obtained either way are consistent and the results from the two runs remain in good agreement with each other, as shown in Table \ref{scale1}.
\begin{table}
\caption{Scale factors for VLA runs 1 (November 2001) and 5 (January 2002). Strict source selection uses only sources with spectra falling more steeply than a spectral index of 0.4 and a correlation coefficient of the spectral index fit greater than 0.8. \label{scale1} (The D(A) configuration had only a very few antennas on the A positions. Almost all the baselines were D-array ones.)}
\begin{tabular}{lcccccc}
\hline
\hline
 VLA          & Cut & Array 	&Field  	& Cal 	& Scale 	& No. \\
 Run   &    &         	& (hrs)  	&            Source       	&  factor		& \\
   &             &	& 	 	&            		      	&  $F_{43}$&	\\
\hline
1		& All  &D   	& 00 		& 3C48       	& $0.72\pm0.04$ 		& 37\\
\hline
5		& All & D(A)	& 15 		& 3C286          	& $0.66\pm0.03$ 		& 54\\ 
\hline
1		& Strict &D  	& 00 		& 3C48       	& $0.68\pm0.03$ 		& 22\\
\hline
5		& Strict & D(A)	& 15 		& 3C286         	& $0.67\pm0.06$ 		& 27\\ 
\hline
\hline
\end{tabular}
\end{table}

Many of the sources are faint at 43\,GHz and were detected at low signal-to-noise in the VLA maps. We can make another cut based on 43\,GHz flux density to include only objects with good detections. To do this we select sources according to the same strict criteria but include only those with predicted 43\,GHz flux densities greater than 20\,mJy. We find $F_{43}$ of $0.68 \pm 0.03$ (from 23 sources). This result is relatively insensitive to the flux density cut-off: increasing the cut to 30\,mJy changes the result to $0.70 \pm 0.04$, with only 10 sources included in the calculation. 

\subsection{Possible reasons for low 43\,GHz measurements}

\begin{table}
\caption{Comparison of our VLA calibrator flux densities with the values from \citealt{pb13} (P\&B) and the effect of the Planck scale factor (\citealt{partridge16})\label{flux_scale}}
\centering
\begin{tabular}{lccccc}
\hline
\hline
Name &           Our	&  P\&B & Ratio & Ratio\\
      	&      value  & 	value  & & \textit{Planck}\\
      	&  		 Jy & 	Jy & & scaling\\
\hline
3C48      		&  	0.53	& 0.62$\pm0.02$&   0.85$\pm0.03$ & 0.80$\pm0.03$\\
3C286       	&  	1.46	& 1.54$\pm0.05$ &  0.95$\pm0.03$ & 0.89$\pm0.03$\\
    \hline
    \hline
\end{tabular}
\end{table}

Several  possibilities present themselves which could explain some of the discrepancy seen in our 43\,GHz flux densities and we briefly examine these here. However, none of the possible effects we describe below is sufficient in itself to explain the 30\% shortfall seen in both VLA runs.

\subsubsection{Overall flux calibration}

The consistency between scale factors in the different VLA runs might suggest that primary calibration is not to blame here, since runs 1 and 5 used different primary calibrators. However, it is possible that the flux density values used for both calibrators---3C48 and 3C286---were in error. 

In 2013 Perley and Butler published a detailed revision of flux density scales in the 1--50~GHz frequency range, based on extensive VLA observations of standard calibrators together with the planet Mars: \citealt{pb13}, hereafter P\&B. Their observations span the years 1983--2012 and the paper provides cubic polynomial coefficients for calculating the spectral flux densities of standard calibrators at a succession of dates. (Whereas 3C286 is extremely stable in time, 3C48 is well known to be variable.) In Table \ref{flux_scale} we show our calibrator flux densities together with the appropriate values derived from P\&B close to our observation dates. Although our values are significantly lower, the discrepancy is clearly not sufficient to account for the whole of our 30\% shortfall.

More recently however, in \citealt{partridge16}, it has been shown that the P\&B flux density scale at 43 GHz may be as much as 6\% too low. Partridge et al. have made coordinated observations with the VLA and the Low Frequency Instrument of the \textit{Planck} satellite of a number of unresolved \textit{Planck} sources.The \textit{Planck} calibration is assumed to be absolute, since it is based on the temperature of the cosmic microwave background. Although at the lower frequencies the difference in the scales is within the error margins, at 43 GHz the P\&B measurements are fainter than those from \textit{Planck} by 6.2\%$\pm1.4$\% which significantly exceeds the uncertainties (P\&B quote 3\%).

We have included the effect of the \textit{Planck} scaling in Table \ref{flux_scale}. It could account for more of the shortfall but still not the whole 30\%.     

\subsubsection{Phase calibration}

 This is especially difficult at high frequency.  VLA runs 1 and 5 were both dominated by D array baselines (run 5 had a few antennas moved to their A array positions but almost all the baselines were D array). Self-calibration was applied to those maps with sufficiently high signal-to-noise ratio. These were typically sources with point-like components having flux densities of about 40~mJy or greater --- approximately 20\% of the sample. Weaker sources which could not be self-calibrated might tend to have depressed flux densities as a result of decorrelation. However, if this were significantly affecting the Bolton04 results we would expect to see a trend in $F_{43}$ with predicted 43\,GHz flux density, and we do not. 

\subsubsection{Pointing errors}

 At higher frequencies the shrinking primary beam of the VLA causes pointing errors to become more significant, since the gain decreases more rapidly with increasing distance from the beam centre. VLA pointing errors can be of the order of 30\,arcsec but, by implementing the method of `referenced pointing' (see e.g. \citealt{rupen97}), they can be reduced to around 5\,arcsec. Since this was  standard practice at the time, it means that, with a primary beam of 1\,arcmin at 43\,GHz, and a possible pointing error of the order of 5 arcsec, any reduction in flux density would have been only about 2\%. To explain a 30\% reduction, the final pointing error would need to be consistently 20-25\,arcsec.
 
\subsubsection{Atmospheric absorption}

At 43\,GHz atmospheric absorption can produce a substantial reduction in the measured flux densities, with the optical depth at certain times in the year reaching about 10\% at zenith (see e.g. \citealt{butler10}). The effect increases with decreasing elevation of the antennae but there is compensation for this if the relevant calibrators are observed at approximately the same elevation as the target sources. This was always the case for the interleaved secondary calibrators, so any flux-density reduction in a target source was offset by a similar reduction in the calibrator. Also, in the calibration of the secondary calibrators from the primary calibrator, the elevations were sufficiently close to avoid serious flux-density discrepancies. Overall, assuming an optical depth at zenith of a maximum of 10\%, we deduce that for the majority of sources any effect was a few percent at most, either up or down. However, for a small group (10 to 15 sources) in run\,5, it is just possible that there could have been a flux-density deficit of as much as 10\% - still considerably less than our 30\% shortfall.

\subsubsection{Elevation gain dependency}

This is another elevation effect which becomes important at high frequencies. It is the change in forward gain of an antenna caused by both the deformation of its surface and the bending of its leg structure. In \citealt{pb13} there are example plots of the normalized gain as a function of elevation for two antennae at 43\,GHz. These show a variation of as much as 10\% or more. However, these effects are known and so were calibrated out using the antenna-based elevation-gain curve tables as part of the standard calibration suite.

\subsubsection{Effect of higher resolution}

 In our original paper, Waldram07, we investigated whether there was any serious `resolving out' of flux at the higher frequencies, owing to the reduction in size of the synthesized beam, and concluded that any effect was insignificant. 

\section{Conclusions}

In this paper we have re-assessed the data, and corresponding source-count predictions, in Waldram07 and Davies13. 

\begin{description}
\item[\textbf{*} ] We present strong evidence that the VLA 43-GHz  flux densities used in our original paper, Waldram07, were substantially underestimated, being on average only about 70\% of their true values. This led to a significant underestimation of the predicted counts at 43, 70 and 90 GHz. For comparison, we have revised these counts using simply the $\alpha_{5,15,22}$ spectral indices and omitting the 43-GHz flux densities. 

\item[\textbf{*} ] Both the revised and original Waldram07 90-GHz counts have been compared with the directly measured count at 95~GHz from the SPT.  At 90~GHz the original count is, as expected, significantly too low.  The revised one, however, is at least 15\% too high, relative to the SPT count. This  indicates that there is indeed some spectral steepening of some sources at frequencies above 22~GHz, even if it is not as extreme as originally assumed.  

\item[\textbf{*} ] We have also re-examined the data used in Davies13 and recalculated the predicted count at 93.2~GHz, after re-calibration at 15.7~GHz. The revised 93-GHz Davies13 count is in good agreement with the SPT count over the range 20-250~mJy.

\item[\textbf{*} ] We have investigated reasons for the problem of the low VLA 43-GHz values and found a number of possible contributory factors, but none is sufficient on its own to account for such a large deficit.  
\end{description} 

Although there is now a directly measured count at 95~GHz, we believe it is important to publish this re-assessment, since a number of authors have made use of our original predictions over a range of frequencies (see, for example,  \citealt{planckXIII}). Also, the good agreement between the counts from Davies13 and the SPT shows that our empirical method of predicting counts works successfully over a well defined range of flux density. It should therefore have other applications where it is difficult to measure high frequency counts directly.

 \section*{Acknowledgments}

We are grateful for a number of useful discussions: with Matthew Davies and Keith Grainge, on the calibration of the AMI data, and with Claire Chandler and Richard Perley, on possible reasons for the low 43-GHz VLA flux density measurements. We thank Gianfranco de Zotti and Marco Tucci for providing their model source counts.

We are also indebted to our reviewer, Bruce Partridge, for his extremely helpful comments.

%\appendix

\label{lastpage}

\end{document}